\documentclass[a4paper,fleqn]{cas-dc}

\usepackage[numbers,sort&compress]{natbib}  
\usepackage{graphicx}    
\usepackage{subcaption}  
\usepackage{amsmath}     

\usepackage{float}       
\usepackage{placeins}    
\usepackage{caption}

\usepackage{multirow} 
\usepackage{booktabs} 

\begin{document}

\let\WriteBookmarks\relax
\def\floatpagepagefraction{1}
\def\textpagefraction{.001}

\shorttitle{Tunable Valley Polarization and Anomalous Hall Conductivity}

\shortauthors{Samiul Islam \textit{et al.}}

\title [mode = title]{Tunable Valley Polarization and Anomalous Hall Effect in Ferrovalley NbX$_2$ and TaX$_2$ (X = S, Se, Te): A First-Principles Study}

\author[1]{Samiul Islam}[
                        orcid=0009-0002-5414-5531]
\ead{samiulislam2616@gmail.com} 
\credit{Conceptualization, Formal analysis, Methodology, Visualization, Software, Investigation, Writing - original draft}

\author[1]{Sharif Mohammad Mominuzzaman}[
                                        ]
\ead{mominuzzaman@gmail.com} 
\credit{Supervision, Resources, Writing-review \& editing}
\author[1]{Ahmed Zubair}[
                        orcid=0000-0002-1833-2244]
\ead{ahmedzubair@buet.ac.bd} 
\credit{Supervision, Conceptualization, Methodology,  Project administration, Resources, Writing - original draft, Writing-review \& editing}
\cormark[1]

\cortext[cor1]{Corresponding author: Ahmed Zubair}

\affiliation[1]{organization={Department of Electrical and Electronic Engineering, Bangladesh University of
Engineering and Technology}, 
                city={Dhaka},
                postcode={1205}, 
                country={Bangladesh}}

\begin{abstract}
Two-dimensional transition metal dichalcogenides lack inversion symmetry and have broken time-reversal symmetry due to the honeycomb structure and intrinsic ferromagnetism, which leads to their valley polarization. Here, we explored the electronic and magnetic properties of the novel ferrovalley materials 1H-NbS$_2$, 1H-NbSe$_2$, 1H-NbTe$_2$, 1H-TaS$_2$, 1H-TaSe$_2$, and 1H-TaTe$_2$ using first-principles calculations based on density functional theory. The materials are dynamically stable bipolar magnetic semiconductors. Among the magnetic semiconductors, NbSe$_2$ showed the maximum Curie temperature of 176.25 K. For these materials, the ferromagnetic state was more favorable than the antiferromagnetic state, indicating robust ferrovalley characteristics. These ferrovalley materials showed a giant tunable valley polarization at K and $\mathrm{K}^\prime$ points in the Brillouin zone without applying any external factors due to intrinsic exchange interactions of transition metal d-orbital electrons and spin-orbit coupling. TaTe$_2$ exhibited an outstanding valley splitting of 541 meV. 
Reversing Bloch electrons' magnetic moment caused an alteration of valley polarization. Additionally, the application of uniaxial and biaxial strain led to the manipulation and variation of the bandgap and valley polarization. Berry curvature exhibited opposite signs and unequal magnitudes at K and $\mathrm{K}^\prime$ points, which led to the anomalous valley Hall effect in these materials. NbS$_2$, NbSe$_2$, and NbTe$_2$ exhibited Berry curvature at unstrained crystals, whereas Berry curvature appeared only in TaSe$_2$ and TaTe$_2$ with the application of strain. These ferrovalley materials exhibited distinct band gaps for spin-up and spin-down electrons, enabling the selective transport of spin-polarized electrons. The bandgap associated with each spin orientation could be modulated through uniaxial or biaxial strain, making them promising for spin filter devices.  Insights from this work will be beneficial for utilizing these ferrovalley materials in advanced valleytronic and spintronic applications such as non-volatile memories, optical and magnetic switches, and valley filters. 
\end{abstract}




\maketitle
\section{Introduction}
Two-dimensional (2D) transition metal dichalcogenides (TMDC) have garnered excellent traction due to their intriguing electronic, electrochemical, thermal, mechanical, magnetic, and optical properties \cite{Akinwande2017else,Jian2021Adv.Opt.Mat,Zubair2022ACS,Huang2021RSC}. Besides, TMDCs exhibit some unique properties like valley contrasting physics \cite{Tony2012Nature}, charge density wave \cite{Zhong2020Nano_research}, and magnetoresistance \cite{Pavel2020App.Sur.sci}. Hence, they have potential applications in superconductors \cite{Xiadong2018NatureCom}, valleytronics, and memory devices \cite{Nguyen2007Nature}. 
In addition, due to the inherent lack of center of symmetry associated with the spin-orbit coupling (SOC) effect from the \(d\) orbitals of the transition metal atoms, they exhibit two degenerate yet inequivalent valley states at the K and $\mathrm{K}^\prime$ points in the hexagonal Brillouin zone (BZ) \cite{Xiong2025RSC,Tony2014NaturePhy}, which is considered a new quantum number called valley degree of freedom (DOF) and acts as a pseudospin. Like charge and spin, valley DOF can be manipulated, which makes them a promising candidate in valleytronic devices, non-volatile storage, valley valves, optical switches, magnetic switches, and valley filters \cite{Beenakker2007NaturePhy,Iwasa2014AmericanAssociation,Lawrence2017APS,Kaustav2023AdvMat}. Besides, the control of valley DOF can be utilized to encode and transmit information \cite{Jie2018Nature}, making them promising candidates for information storage as the significant separation between inequivalent valleys in momentum space reduces intervalley scattering from deformation and long wavelength phonons \cite{Xiao2018Wiley,Xiaodong2016NatureReview}. This surpasses modern charge-based semiconductor technologies \cite{Xiaodong2016NatureReview}. Recent studies demonstrated that materials like monolayer TMDCs exhibit robust valley polarization, which can be controlled through external stimuli, such as electric fields or mechanical strain, further supporting their potential in advanced information processing technologies \cite{xiangfeng2018RSC,Jens2024Nature,Kaustav2023AdvMat}.

The concept of 2D valleytronics originated in graphene \cite{Qian2007APS}. In materials with time-reversal symmetry (TRS), the electronic states at the inequivalent K and $\mathrm{K}^\prime$ points are degenerate, as the symmetry enforces identical energy levels, preventing spontaneous valley polarization. Breaking the valley degeneracy in these materials is crucial, as it would allow the control of valley polarization, which is essential for memory and logic applications \cite{Xiaodong2012PhysicsReview,Lei2017Nanoscale}, in next-generation electronics \cite{Qian2007APS,Xiaodong2012PhysicsReview}, and optoelectronics \cite{Iwasa2014AmericanAssociation}. Valley-dependent electronics and optoelectronics based on semi-metallic graphene, a representative 2D material, were theoretically proposed \cite{Qian2007APS,Xiaodong2012PhysicsReview}. In 2D materials, valley polarization, the selective population of electrons in one of the inequivalent K or $\mathrm{K}^\prime$ valleys in the BZ, depends critically on the presence or absence of TRS and inversion symmetry. When TRS is preserved, the energies at K and $\mathrm{K}^\prime$ remain degenerate, making valley polarization challenging. However, breaking TRS lifts this degeneracy by shifting the energy of each valley differently, creating an effective magnetic field \( B_{\text{eff}} \) that results in a valley-specific energy difference \(\Delta E = g \mu_B B_{\text{eff}}\), where \( g \) is the Landé g-factor and \( \mu_B \) is the Bohr magnetron. However, the valley energy is degenerate owing to TRS in most 2D TMDC  materials,  which restricts their practical applicability.

Breaking TRS between K and $\mathrm{K}^\prime$ valleys can be achieved by using external magnetic fields \cite{Potemski2016APSphyreview,Yihang2017NatureNano}, magnetic proximity effects \cite{Qiming2019APSphyreview,Wei2020APSphyreview}, optical Stark effect \cite{Tony2017Nature}, optical excitation \cite{Atac2015Nature}, magnetic doping \cite{Li2020APS}
, or electric fields \cite{Ying2021RSC}. However, these external methods have limitations, as valley polarization disappears once the external field is removed, which is unsuitable for non-volatile valleytronic devices. Additionally, the effectiveness of magnetic-field-induced valley polarization is low (0.1–0.2 meV/T) \cite{Yin2019RSC}, and magnetic substrates can impact device performance by increasing dimensions, causing minor Zeeman splitting, and altering band structures \cite{Yin2019RSC}. Magnetic dopants tend to cluster on TMDC surfaces, leading to carrier scattering. At the same time, circularly polarized light, though adequate for valley polarization in TMDC, requires complex tuning and has a short luminescence lifetime \cite{Jiwoong2014AmericanAssociation}.

Ferrovalley (FV) materials, such as the 2H-VSe\(_2\) monolayer \cite{Xiangang2016NatureCom}, offer an alternative by achieving spontaneous valley polarization through intrinsic ferromagnetism, benefiting from SOC and intrinsic exchange interactions. This spontaneous polarization remains stable without external fields, which is ideal for valleytronic and spintronic devices. Few FV materials have been discovered, including 2H-VSe\(_2\) \cite{Xiangang2016NatureCom}, VAgP\(_2\)Se\(_6\) \cite{Jing2018RSC}, LaBr\(_2\) \cite{Ying2019AIP}, and GdX\(_2\) (X = Cl, F) \cite{Yong2021WileyPhysicaStatus}, GdI$_2$ \cite{Feng2021APS}, ScX$_2$ (X= Br, Cl, I) \cite{Xianmin2023Acta}, FeCl$_2$ \cite{Yandong2022ChemPhy}, YX$_2$ (X= Br, Cl, I) \cite{WeiBing2023APS} which combine spin polarization and valley properties, making them promising for non-volatile applications. Previous studies identified significant challenges in valleytronic devices, primarily due to the difficulty in achieving spontaneous valley polarization, clustering of magnetic atoms, and the complexity of the mechanism to achieve valley polarization. Though many FV materials were reported to exhibit spontaneous valley polarization, the valley splitting energy was low, which hindered the practical realization of valleytronic devices. Among these, the most prominent valley polarization was reported for GdI$_2$, which was 149 meV \cite{Feng2021APS}.

In this study, we explored a novel family of 2D FV materials that exhibited ease of exfoliation and demonstrated spontaneous valley polarization with significantly enhanced valley polarization energy. This elevated energy surpassed the threshold of electronic noise under ambient and extreme conditions, thereby offering promising potential for practical valleytronic devices. Moreover, we investigated the tunability of valley-splitting energy in these materials. We reported the identification of a new class of FV materials, including 1H-NbS$_2$, 1H-NbSe$_2$, 1H-NbTe$_2$, 1H-TaS$_2$, 1H-TaSe$_2$, and 1H-TaTe$_2$. While NbS$_2$ and NbSe$_2$ were previously studied, we provided a comparative analysis of these materials alongside other members of the same group. Our findings revealed unique insights into the behavior of group V TMDCs, highlighting their potential for advanced valleytronic technologies. They exhibited intrinsic valley polarization without external manipulation. Unlike conventional approaches where valley polarization is induced through external fields or magnetic doping, these materials demonstrated robust spontaneous valley polarization driven by their intrinsic magnetic properties. These materials showed superior valley polarization energy. 
We also demonstrated the manipulability of the materials through magnetic moment reversal and uniaxial and biaxial strain. To show the valley-contrasting behaviors of the TMDCs, we calculated the quantum transport properties like Berry curvature and anomalous Hall conductivity (AHC), which confirmed the potential of these TMDCs for valleytronic and spintronic applications.

\section{Computational Details}
First-principles calculations were performed using density functional theory (DFT) using the Quantum ESPRESSO package \cite{quantum_espresso_1}. The projector augmented wave (PAW) method, coupled with the Perdew-Burke-Ernzerhof (PBE) exchange-correlation functional, was used to model the electron-ion interactions. Unlike norm-conserving pseudopotentials, PAW pseudopotentials offer higher accuracy in describing localized orbitals, especially for transition metals with significant SOC \cite{Zubair2022ACS}. For transition metal atoms, \(d\) orbitals were treated as valence electrons, whereas \(d\) orbitals were considered as core electrons for chalcogen atoms in the pseudopotentials. The Hubbard U  parameter was applied to account for strong on-site Coulomb interactions in the \(d\) orbitals\,\cite{hubbard_param}. The Hubbard parameters were obtained using the linear response approach implemented in Quantum ESPRESSO. This method calculates the Hubbard U by perturbing the on-site occupation of localized orbitals and measuring the response in the electronic structure self-consistently. The optimized Hubbard parameters for Nb and Ta atoms' \(d\) orbitals for NbS$_2$, NbSe$_2$, NbTe$_2$, TaS$_2$, TaSe$_2$, and TaTe$_2$ were 3.1148, 3.1231, 3.0448, 2.3392, 2.5686, and 2.1982, respectively. Cold smearing, with a width of 0.005 Ry, enhanced convergence.

In our study, the unit cell of a TMDC with hexagonal symmetry, belonging to the space group P63/mmc, was utilized. The structure featured trigonal prismatic coordination of the transition metal layer sandwiched between two chalcogen layers. This arrangement formed a periodic honeycomb-like lattice characteristic of monolayer TMDCs. A 15 \AA{} vacuum gap along the z-axis minimized interlayer interactions. A kinetic energy cutoff of 55 Ry was applied to define the maximum plane-wave energy used in expanding the electronic wavefunctions, and a charge density cutoff of 550 Ry was employed to specify the resolution of the charge density grid. Variable cell relaxation was performed on the unit cell, allowing both atomic positions and lattice parameters to adjust. The relaxation process continued until the total energy converged to a threshold of \(10^{-8}\) Ry per atom, and the forces between atoms were reduced below \(10^{-5}\) Ry/Bohr. Relaxation and self-consistent field (SCF) calculation were performed with a (24 $\times$ 24 $\times$ 1) \(\Gamma\)-centered Monkhorst-Pack grid for BZ sampling and a denser (48 $\times$ 48 $\times$ 1) k-point grid for density of states (DOS) calculations. To assess the energetic stability of the system, we utilized the total ground-state energy derived from self-consistent field (SCF) calculations. This approach ensures accurate energy evaluation by accounting for electron-electron interactions through a self-consistent solution. We utilized the Heyd–Scuseria–Ernzerhof (HSE06) hybrid functional within Quantum ESPRESSO to compute the electronic bandgap. Hybrid functionals like HSE06 combine a portion of exact Hartree–Fock exchange with density functional theory (DFT) exchange-correlation, addressing the typical bandgap underestimation in standard DFT methods. We performed Bader charge analysis using the charge density obtained from SCF calculations in Quantum ESPRESSO. This analysis was conducted using the Bader code developed by the Henkelman group \cite{bader}. Additionally, we evaluated the average electrostatic potential by summing the bare ionic potential and the Hartree potential derived from the same SCF calculations. For evaluating antiferromagnetic (AFM) and ferromagnetic (FM) properties for TMDCs, we used a (2 $\times$ 2 $\times$ 1) supercell after variable cell relaxation.

\begin{figure*}[b!]
    \centering
    \includegraphics[width=1.0\textwidth]{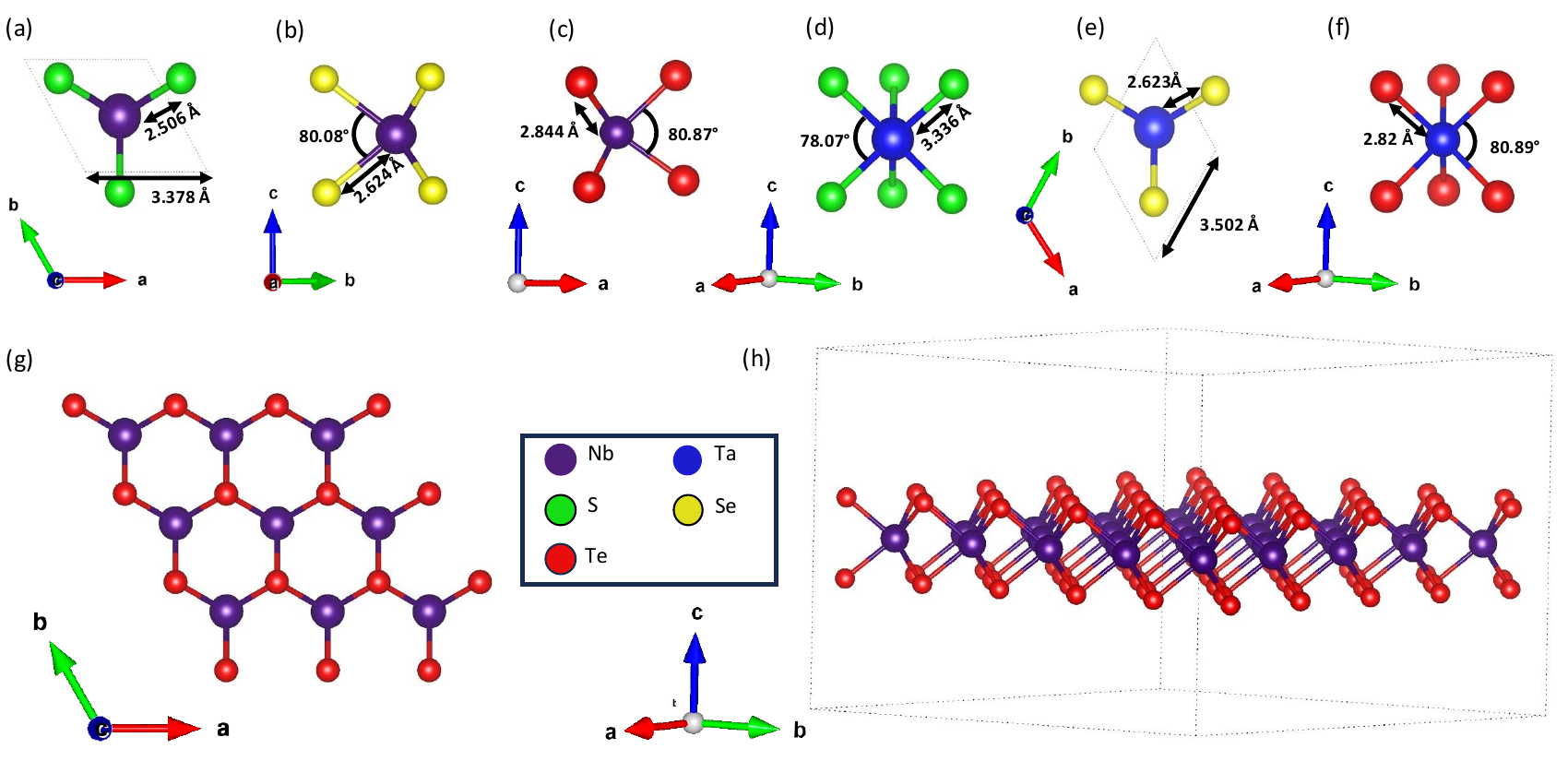}
    \caption{Optimized structures of different monolayers: (a) NbS$_2$ (top view), (b) NbSe$_2$ (side view), (c) NbTe$_2$ (side view), (d) TaS$_2$ (side view), (e) TaSe$_2$ (top view), (f) TaTe$_2$ (side view), (g) top view of NbTe$_2$ (3 $\times$ 3 $\times$ 1) supercell, (h) side view of NbTe$_2$ (5 $\times$ 5 $\times$ 1) supercell. Purple, blue, green, yellow, and red atoms represent Nb, Ta, S, Se, and Te, respectively.}
    \label{fig:optimized_parameters}
\end{figure*}

To investigate valley-related properties, we incorporated SOC effects using fully relativistic PAW pseudopotentials. We computed the Berry curvature and anomalous Hall conductivity (AHC) employing Wannier90 \cite{wannier90} and WannierTools \cite{wannier_tools}. A (13 $\times$ 13 $\times$ 1) non-self-consistent k-point grid was utilized to generate maximally localized Wannier functions, facilitating accurate interpolation of the electronic structure. The non-self-consistent calculation to obtain Wannier functions was performed in Quantum ESPRESSO using fully relativistic pseudopotentials.

Dynamic stability was verified using a (2 $\times$ 2 $\times$ 1) supercell through the finite displacement method implemented in Phonopy \cite{phonopy}. This approach systematically displaces atoms within the supercell and calculates the resulting forces to construct the dynamical matrix from which phonon frequencies are determined. To obtain accurate forces, stresses, and ground-state energies, SCF calculations for the displaced atomic configurations were performed using Quantum ESPRESSO.

\section{Results and Discussion}
The exfoliation of 2D materials has progressed significantly with techniques like liquid-phase exfoliation (LPE). It provides a scalable and cost-effective approach for producing large quantities of solution-processed 2D materials, making it particularly suitable for electrochemical applications. Notably, niobium diselenide (NbSe$_2$) was successfully exfoliated using a two-step vapor deposition method, as well as LPE \cite{Sofia2024MDPI}. Likewise, thin films of NbS$_2$ \cite{Yang2019ACS}, TaSe$_2$ \cite{Hua2013Wiley}, and metallic telluride nanosheets (e.g., NbTe$_2$ and TaTe$_2$) were effectively produced through exfoliation techniques \cite{Zhang2018WileyAdvMat}. These materials, which can be readily synthesized, exhibit hexagonal honeycomb lattice structures with broken inversion symmetry, a structural attribute that endows them with distinctive electronic properties.

\subsection{Structural Properties and Dynamic Stability}

We optimized unit cells of pristine 1H-NbS$_2$, 1H-NbSe$_2$, 1H-NbTe$_2$, 1H-TaS$_2$, 1H-TaSe$_2$, 1H-TaTe$_2$ by relaxing them until the convergence threshold mentioned were achieved. To avoid interlayer interaction, we added a vacuum level of 15 \AA\ normal to the 2D sheet. Relaxed pristine TMDCs had hexagonal symmetry. The transition metal was sandwiched between two layers of chalcogens, forming a trigonal prismatic coordinated with six chalcogen atoms. The optimized lattice parameters (a=b) for NbS$_2$ and NbSe$_2$ were 3.378 and  3.479 \AA{}, respectively, which is comparable with the previous report \cite{Ying2021NanoReas}. Lattice parameters of NbTe$_2$, TaS$_2$, TaSe$_2$, TaTe$_2$ were found to be 3.699, 3.336, 3.502, and 3.705 \AA{}, respectively. In niobium chalcogenides, the Nb-X (X = S, Se, Te) bond length increased with the atomic radius of the chalcogen. Nb-S bond length was 2.506 \AA, the Nb-Se bond length was about 2.624 \AA, and the Nb-Te bond length was around 2.844 \AA. This trend reflected the increasing size of the chalcogen atoms from sulfur to tellurium, leading to longer Nb-X bonds. For Ta atoms, the bond lengths showed a similar increasing length with S, Se, and Te, where bond lengths were 2.480, 2.623, and 2.820 \AA{}, respectively. The bond angles within TMDCs exhibited a systematic increase with the atomic number of the chalcogen atom. This trend arose due to the increasing atomic radius and decreasing electronegativity from sulfur (S) to selenium (Se) to tellurium (Te), which influenced the spatial configuration of the crystal lattice. For niobium-based TMDCs, the chalcogen–metal–chalcogen bond angles increased. The bond angle in NbS$_2$ is 77.79$^\circ$, which was a maximum of 80.87$^\circ$ in NbTe$_2$. A similar trend was observed in tantalum-based TMDCs, where the bond angles were 78.07$^\circ$ for TaS$_2$, 79.47$^\circ$ for TaSe$_2$, and 80.89$^\circ$ for TaTe$_2$. The optimized lattice parameters, bond lengths, and bond angles are presented in Table \ref{table:combined}.

\begin{table*}[t] 
\centering
\caption{Structural, formation energy, work function, and Curie temperature of pristine TMDCs}
\label{table:combined}
\resizebox{\textwidth}{!}{ 
\begin{tabular}{lcccccc}
\toprule
\textbf{Crystals} & \textbf{Lattice parameter (\AA)} & \textbf{Bond length (\AA)} & \textbf{Bond angle ($^\circ$)} & \textbf{Formation energy (eV)} & \textbf{Work function (eV)} & \textbf{Curie Temperature (K)} \\
& & & $\angle$ chalcogen-metal-chalcogen & (per atom) & & \\
\midrule
NbS$_2$  & 3.378 & 2.506 & 77.79  & -4.7841  & 5.67  & 161.73 \\
NbSe$_2$ & 3.479 & 2.624 & 80.08  & -4.26322 & 5.33  & 176.25 \\
NbTe$_2$ & 3.699 & 2.844 & 80.87  & -3.6988  & 4.82  & 154.21 \\
TaS$_2$  & 3.336 & 2.480 & 78.07  & -5.4565  & --    & --    \\
TaSe$_2$ & 3.502 & 2.623 & 79.47  & -4.8352  & 5.11  & 79.17  \\
TaTe$_2$ & 3.705 & 2.820 & 80.89  & -4.2861  & 4.75  & 79.27  \\
\bottomrule
\end{tabular}
}
\end{table*}

With the increasing atomic number in the same period in the periodic table, the lattice parameter and the bond length showed an increasing trend. This was due to the larger size of atoms, with the increasing atomic number having the same period in the periodic table. The bond angle showed the same trend. With increasing size, the bond angle increased. The optimized crystals are illustrated in Fig.\,\ref{fig:optimized_parameters}.

We predicted the energetic stability from the formation energy of the TMDCs. To calculate formation energy, we used the ground state energy and the single-atom ground state energy obtained from the SCF calculations. The formation energy of the alloy was determined by,

\begin{equation}
E_{\text{formation}} = E_{\text{M}\text{X}_{2}} - E_{\text{M$_{total}$}} - 2 \times E_{\text{X$_{total}$}}.
\end{equation}
Here, $E_{\text{formation}}$ was the ground state total energy of the pristine crystals and $E_{\text{M}}, E_{\text{X}}$ were the single atom energy of the atoms. Here, M = (Nb, Ta) and X = (S, Se, Te).

The negative formation energy of the TMDCs suggests that energy will be released during the formation of the crystals. This process is an exothermic process. The largest formation energy was exhibited by TaS$_2$, which was -5.457 eV per atom. The formation energy was in the range of around 4 eV per atom, which is comparable to other crystals formation energy. The formation energies are reported in Table \ref{table:combined}.

\begin{figure*}[ht]
    \centering
    \includegraphics[width=\textwidth]{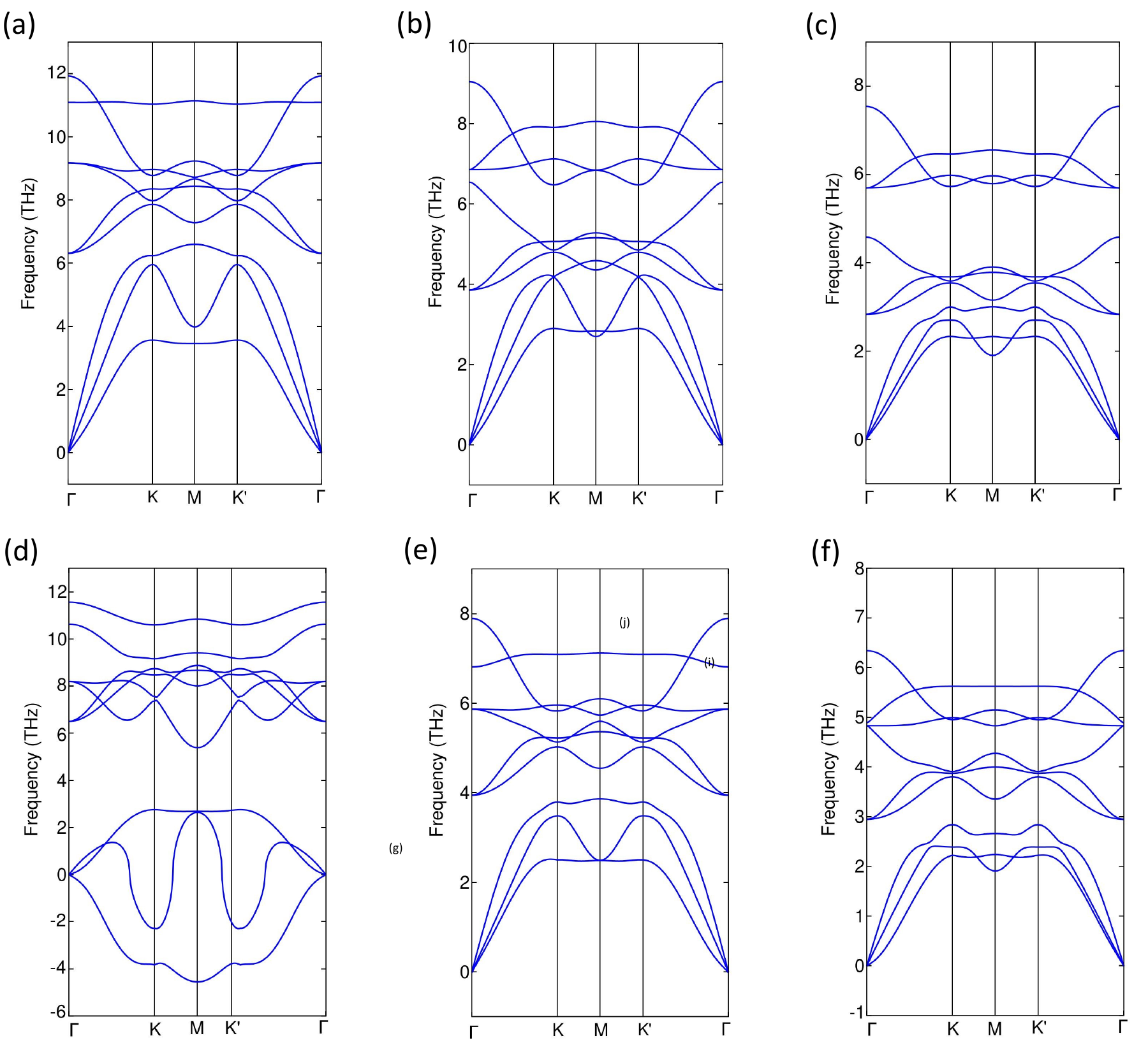} 
    \caption{Phonon spectra of (a) NbS$_2$, (b) NbSe$_2$, (c) NbTe$_2$, (d) TaS$_2$, (e) TaSe$_2$, (f) TaTe$_2$.}
    \label{fig: dynamic_stability_TMDCs}
\end{figure*}

We calculated the dynamic stability of the crystals by calculating the phonon dispersion. The finite displacement method was implemented to calculate the phonon dispersion, where 6 supercells were created, displacing the atomic positions. Then, the phonon dispersion relation was evaluated from the force and stress on the system. TaS$_2$ showed the most significant release of energy during formation; however, TaS$_2$ was found to be dynamically unstable due to negative frequency in the phonon spectra. Except for TaS$_2$, all other crystals were dynamically stable.  The phonon dispersion relationship is in Fig.\,\ref{fig: dynamic_stability_TMDCs}.

The phonon dispersion relation showed that NbS$_2$, NbSe$_2$, NbTe$_2$, TaSe$_2$, and TaTe$_2$ had no imaginary frequency, which indicated the dynamic stability of the crystals. There were three acoustic and six optical phonon branches present in the crystal. Except for NbS$_2$ and NbSe$_2$, there was a phonon gap between acoustic and optical phonons for these crystals. A more significant phonon gap can reduce phonon-phonon scattering, potentially increasing thermal conductivity. TaS$_2$ exhibits a significant imaginary (negative) frequency in the dispersion graph. TaS$_2$ is not stable at room temperature. Hence, we discarded further discussion on TaS$_2$. \\

\subsection{Bader Charge Analysis and Electrostatic Potential}
The bond nature in the crystals was predicted through Bader charge analysis. In these TMDC crystals, the oxidation state of Nb and Ta is +4. However, Bader charge analysis revealed that the actual charge transfer by Nb and Ta atoms was +1.661e, +1.451e, +1.137e, +1.529e, and +1.120e, respectively, for NbS$_2$, NbSe$_2$, NbTe$_2$, TaSe$_2$, and TaTe$_2$. This result suggested that the transition metals carried a partial positive charge. While the chalcogen atoms (S, Se, and Te) held partial negative charges, indicating electron transfer between the metal and chalcogen atoms. Such a distribution of charges highlighted a partial ionic character in the covalent bonding between atoms. NbS$_2$ showed the most ionic characteristics. Table \ref{table:bader_charge_analysis} illustrates the Bader charge analysis.

\begin{table}[h]
\centering
\small 
\caption{Bader charge analysis of TMDCs}
\label{table:bader_charge_analysis}
\begin{tabular}{lcc}
\hline
\textbf{Crystals} & \textbf{Transition metal} & \textbf{Chalcogen} \\
 & \textbf{charge (e)} & \textbf{charge (e)} \\
\hline
NbS$_2$ & +1.661 & -0.832 \\
NbSe$_2$ & +1.451 & -0.727 \\
NbTe$_2$ & +1.137 & -0.563 \\
TaSe$_2$ & +1.529 & -0.764 \\
TaTe$_2$ & +1.120 & -0.557 \\
\hline
\end{tabular}
\end{table}

\begin{figure*}[ht]
    \centering
    \includegraphics[width=\textwidth]{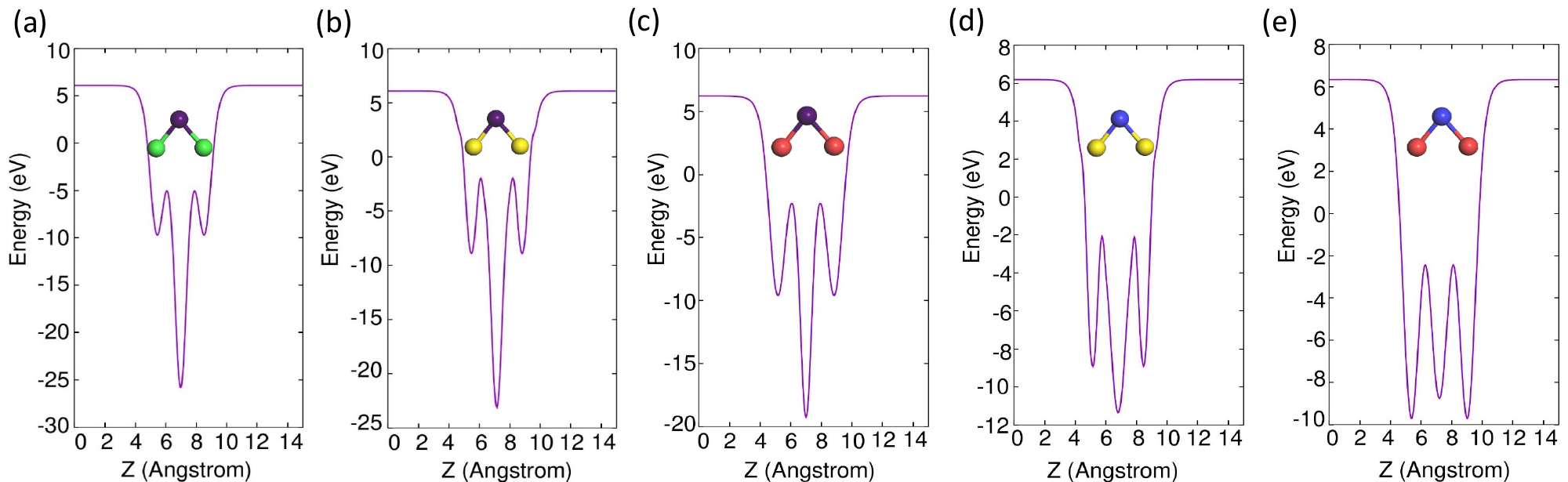} 
    \caption{Average electrostatic potential of TMDC structures: (a) NbS$_2$, (b) NbSe$_2$, (c) NbTe$_2$, (d) TaS$_2$, (e) TaSe$_2$, (f) TaTe$_2$. Purple, blue, green, yellow, and red atoms are Nb, Ta, S, Se, and Te, respectively.}
    \label{fig: work_function}
\end{figure*}

The work function ($\phi$) of TMDCs is a crucial parameter that impacts their electronic properties, surface reactivity, and applicability in electronic and optoelectronic devices. The average electrostatic potential is plotted in Fig.\,\ref{fig: work_function}. The oscillations in the average electrostatic potential were more negative for transition metal atoms (Nb, Ta) than chalcogen atoms (S, Se, and Te) due to differences in electronic configuration, electronegativity, and charge distribution. Transition metals have highly localized \(d\) orbitals and higher local charge densities, resulting in deeper potential wells. In contrast, chalcogen atoms, with more diffuse \(p\) orbitals and higher electronegativity, attracted some electron density, leading to smoother and less negative oscillations. Additionally, the compact atomic radii of transition metals enhanced the nuclear potential, whereas the larger, more diffuse chalcogen orbitals resulted in a less pronounced potential. This interplay of orbital localization and charge redistribution was critical for understanding the electronic properties of TMDCs. From the average electropotential, we calculated the work function. The work function was calculated by,
\begin{equation}
\phi = E_{\text{vacuum}} - E_{\text{Fermi}}.
\end{equation}
The work function is presented in Table\,\ref{table:combined}. NbS$_2$ had a relatively high work function of approximately 5.67 eV, suggesting enhanced surface stability and a reduced tendency for electron emission. On the other hand, NbTe$_2$ and TaTe$_2$, with work functions around 4.82 and 4.75 eV, respectively, may be more suitable for applications that require lower electron emission barriers. These values align with typical TMDC materials such as MoS$_2$ \cite{Ganesh2021ACS}, whose work functions range between 4.50 and 5.00 eV. Factors such as the specific transition metal and chalcogen atoms, layer thickness, and structural variations influence the work function and overall electronic behavior of TMDCs \cite{Joon2021APS}. Understanding the work functions of these TMDCs is crucial for optimizing their performance in applications such as field-effect transistors and catalytic processes, where control over electron emission and surface interactions is essential.

\subsection{Magnetic Properties}
\begin{figure}[ht]
    \centering
    \includegraphics[width=\columnwidth]{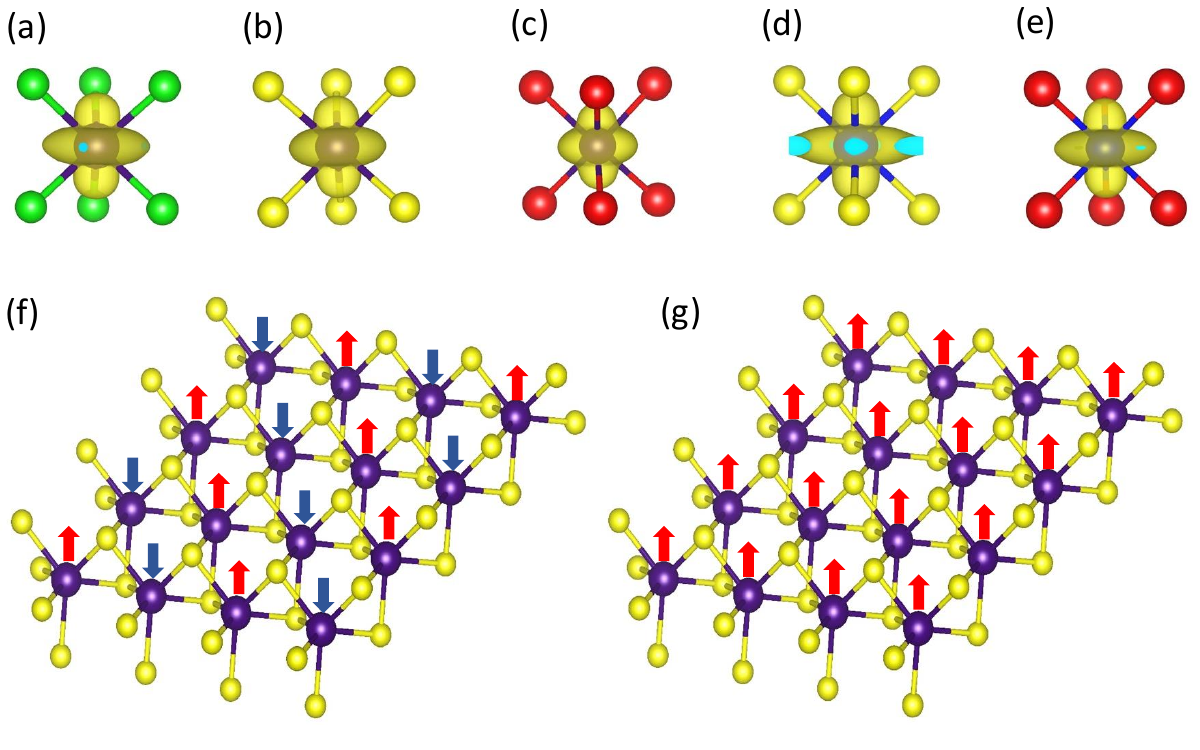}
    \caption{Spin polarization density of the TMDCs (a) NbS$_2$ (b) NbSe$_2$ (c) NbTe$_2$ (d) TaSe$_2$ (e) TaTe$_2$. The yellow region indicates the spin polarization density. Direction of atomic magnetic moments for NbSe$_2$ (f) AFM configuration (g) FM configuration.}
    \label{fig: spin_density}
\end{figure}

The spatial distribution of spin density difference can predict the magnetism in a material. Uniform spin density indicates ferromagnetic ordering, while alternating regions of spin density suggest antiferromagnetic ordering. Spin polarization density is given by,
\begin{equation}
\Delta \rho_{\text{spin}} = \rho_{\text{spin}\uparrow} - \rho_{\text{spin}\downarrow}.
\end{equation}
Fig.\,\ref{fig: spin_density} elucidates the spin polarization density of the TMDCs. The yellow isosurface represents regions of high spin density, indicating areas where there is a significant spin polarization. The specific structure shown suggested a central atom having an unpaired electron in an orbital. The spin polarization in the materials was due to only one spin. No opposite spin is seen in the spin polarization plots. The central atom likely had a significant magnetic moment, contributing to the observed spin density. This predicted the materials to be ferromagnetic.

To confirm the FV character of the TMDCs, we calculated the FM energy and AFM energy of the monolayers. The most favorable configuration has the lowest ground state energy. The FM ground state and AFM ground state energy predicted the magnetic ground state for a (2 $\times$ 2 $\times$ 1) supercell. Figure\,\ref{fig: spin_density} illustrates the FM and AFM configurations. For AFM and FM calculations, the AFM and the FM supercells were optimized by variable cell relaxation, and the ground state energy was achieved using the SCF calculation.  The FM and AFM ground state energies indicated that the FM state was more favorable than the AFM state, as the FM ground state energies were lower than the AFM ground state energies. The magnetic ground state energy is presented in Table \ref{table:magnetic_ground_state_energies}.

\begin{table}[H] 
\centering
\caption{Magnetic Ground State Energies of TMDCs}
\label{table:magnetic_ground_state_energies}
\resizebox{\linewidth}{!}{ 
\begin{tabular}{lccc}
\hline
\textbf{Crystal} & \textbf{FM energy (eV)} & \textbf{AFM energy (eV)} & \textbf{Energy difference} \\
 & & & $E_{FM} - E_{AFM}$ (eV) \\
\hline
NbS$_2$  & -25406.0866 & -25405.5855 & -0.5011 \\
NbSe$_2$ & -38795.4162 & -38794.8701 & -0.5461 \\
NbTe$_2$ & -57651.5818 & -57651.1040 & -0.4778 \\
TaSe$_2$ & -60178.2764 & -60178.0311 & -0.2453 \\
TaTe$_2$ & -79034.6263 & -79034.3807 & -0.0181 \\
\hline
\end{tabular}
}
\end{table}

The FM and AFM energies confirmed that NbX$_2$ and TaX$_2$ were FM in the ground state. The magnetic moment of the TMDCs was calculated, and the FM materials showed a total magnetization of 1 Bohr magneton per unit cell.

For 2D materials, Monte Carlo (MC) simulations of the Heisenberg model are effective in accurately estimating transition temperatures \cite{Thomas2018IOP}. The Curie temperature, \(T_C\) was determined using an equation derived from MC simulations of the XY model, which is well-suited for capturing critical behavior in 2D magnetic systems.
\begin{equation}
T_C = \frac{0.89}{8k_B}(E_{AFM} - E_{FM}),
\end{equation}
here, $k_B$ is the Boltzmann constant, and $E_{FM}$ and $E_{AFM}$ were the energies of FM and AFM configurations. The $T_C$ was calculated and presented in Table \ref{table:combined}. The Curie temperature for NbSe$_2$ was the largest, which was found to be 176.25 K and comparable with the previous report \cite{Yan2023AIP}. TaSe$_2$ had the lowest Curie temperature, 79.17 K. However, $T_C$ values for all studied material were below the room temperature.

\subsection{Electronic Properties}
 The band structure was plotted to determine the electronic characteristics of the TMDCs. The band structure without SOC predicted that all the TMDCs- NbS$_2$, NbSe$_2$, NbTe$_2$, TaSe$_2$, and TaTe$_2$ were semiconductors with a decent bandgap. The bandgaps of the TMDCs are 0.5818, 0.7911, 0.7700, 0.3977, and 0.1947, respectively. 
 The calculated bandgap of NbS$_2$ is 0.2188 eV, which aligns well with previously reported values \cite{Ying2021NanoReas}. However, our computed bandgap for NbSe$_2$ shows a discrepancy compared to existing literature, which may be attributed to differences in computational methods, such as the choice of exchange-correlation functionals or the inclusion of spin-orbit coupling effects. The bandgap is given in Table \ref{table:merged_TMDCs}. Due to magnetic exchange interactions, semiconductors exhibited spin splitting between spin-up and spin-down electrons. The conduction band minimum and valence band maximum were spin-polarized in these materials, resulting in distinct electron and hole transport spin channels. This unique property classified them as bipolar magnetic semiconductors (BMS)~\cite{Xingxing2024APS}.

\begin{figure*}[Ht]
    \centering
    \includegraphics[width=\textwidth]{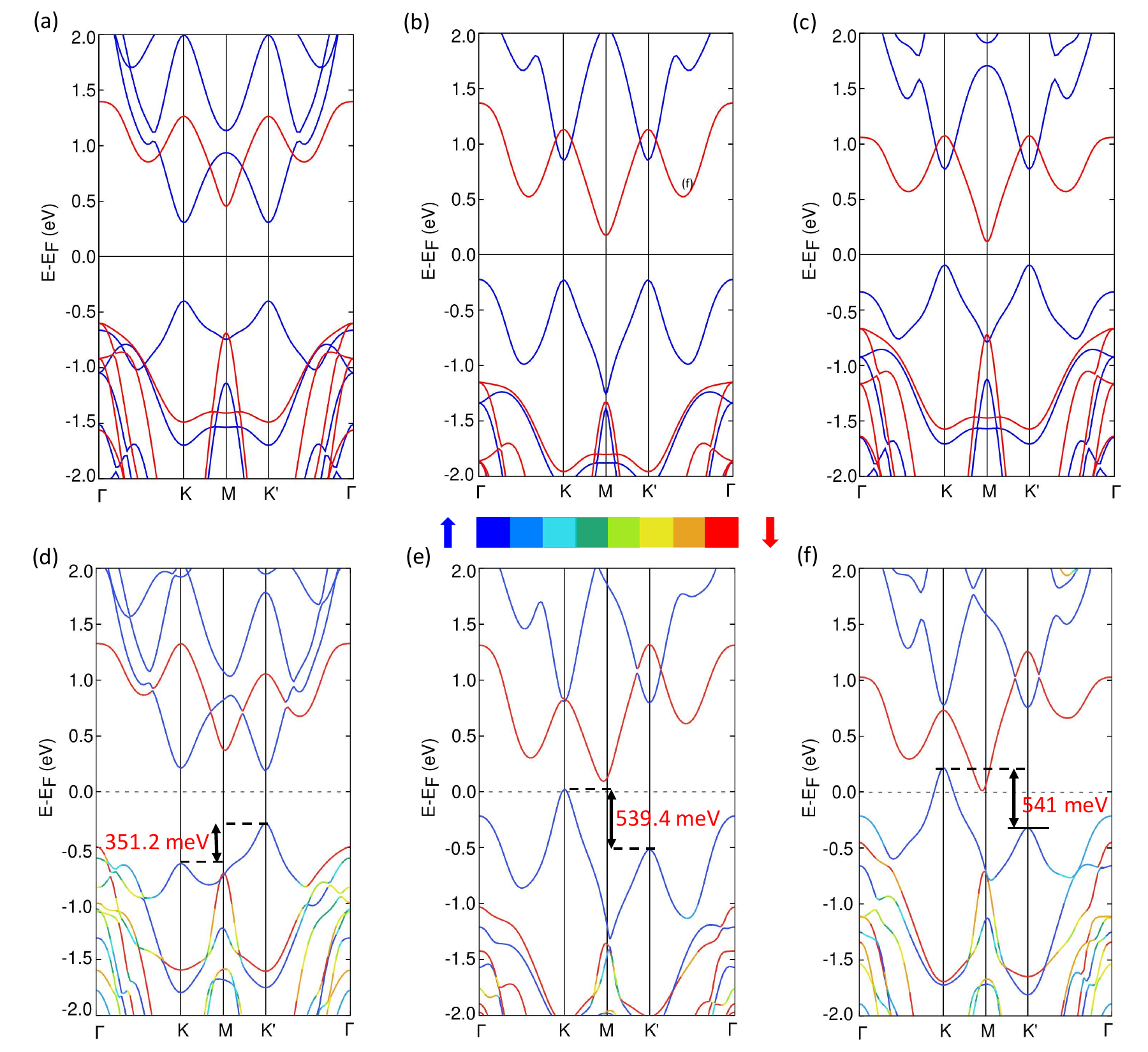} 
    \caption{Spin-polarized band structure for TMDCs without SOC: (a) NbTe$_2$, (b) TaSe$_2$, (c) TaTe$_2$. Spin-up and spin-down states are distinguished by blue and red lines, respectively, to visualize spin polarization. Band diagram incorporating SOC for (d) NbTe$_2$, (e) TaSe$_2$, (f) TaTe$_2$. The color gradient from blue to red depicts expectation values of the spin operator along the [001] direction, ranging from $+\frac{1}{2}$ to $-\frac{1}{2}$.}
    \label{fig: band_structure_TMDCs}
\end{figure*}
\begin{figure*}[ht]
    \centering
    \includegraphics[width=\textwidth]{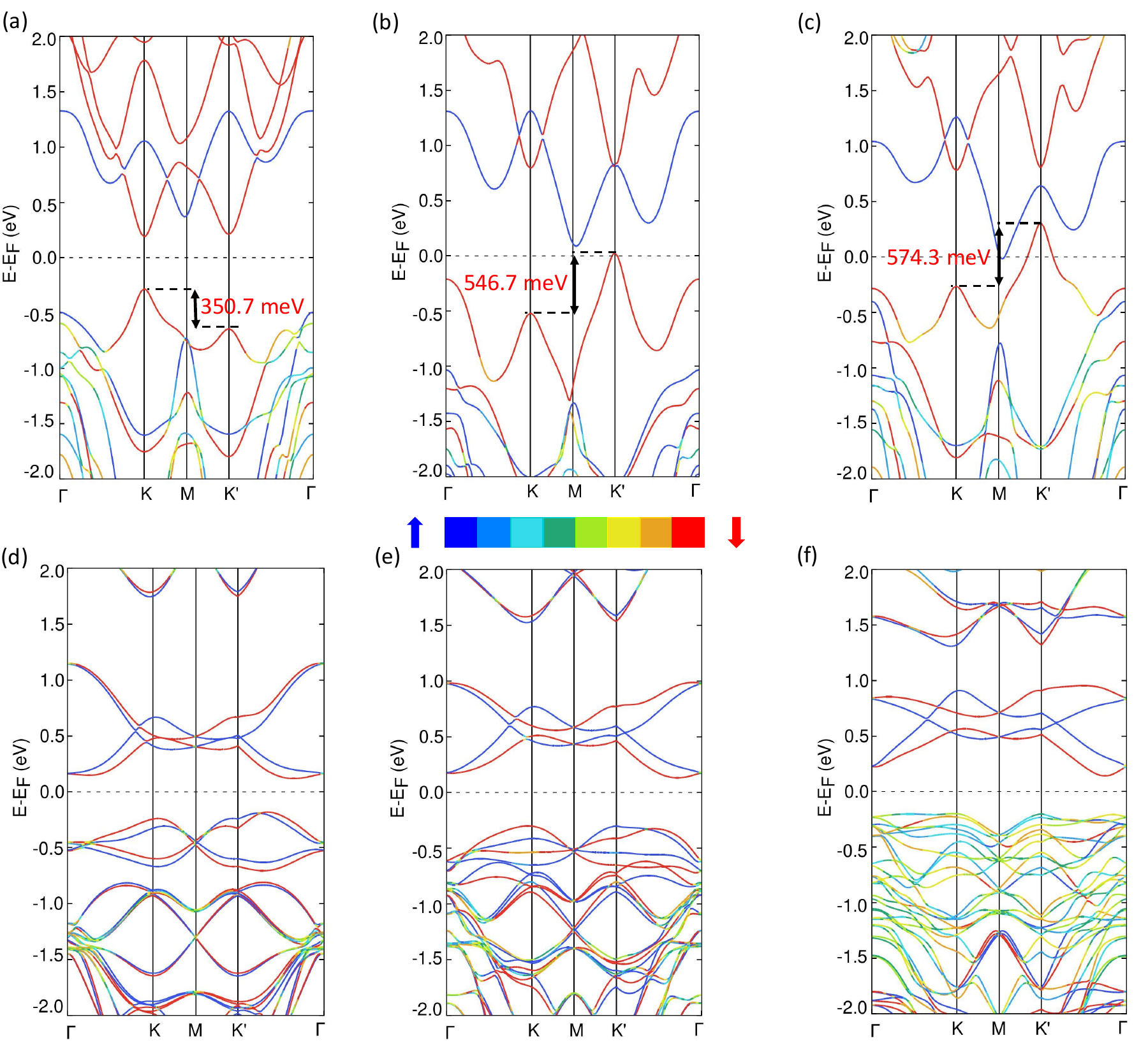} 
    \caption{Band structure after reversal of magnetic moment with SOC for different TMDCs: (a) NbTe$_2$, (b) TaSe$_2$, (c) TaTe$_2$. Antiferromagnetic band structure with SOC of different TMDCs: (d) NbS$_2$, (e) NbSe$_2$, (f) NbTe$_2$. The color gradient from blue to red depicts expectation values of the spin operator along the [001] direction, ranging from $+\frac{1}{2}$ to $-\frac{1}{2}$.}
    \label{fig: manipulation}
\end{figure*}

The monolayer semiconductors NbS$_2$, NbSe$_2$, NbTe$_2$, TaSe$_2$, and TaTe$_2$ exhibited promising spin filter properties, characterized by non-degenerate spin-up and spin-down energy states. The spin-up channels displayed bandgaps of approximately 0.8227, 1.0254, 0.8556, 1.0759, and 0.9159 eV, respectively. In contrast, the spin-down channels exhibited bandgaps of 1.5359, 1.3915, 0.9279, 1.3269, and 0.7291 eV, respectively. These bandgap values, as illustrated in Table \ref{table:merged_TMDCs}, and given in Fig.\,\ref{fig: band_structure_TMDCs} and the Supplementary Material, highlight the potential for spin-selective transport, where each spin channel can be independently controlled, a desirable property for spintronics applications. In BMS, fully spin-polarized currents with adjustable spin polarization can be generated and modulated by applying a gate voltage \cite{Xingxing2024APS}. This characteristic allows for precise control over spin-dependent transport, making BMS materials highly promising for spintronic applications where the direction of spin polarization can be reversibly tuned \cite{Xingxing2024APS}. BMS like NbS$_2$, NbSe$_2$, NbTe$_2$, TaSe$_2$, and TaTe$_2$ demonstrated spin-filtering capabilities, making them suitable for a range of spintronic applications. These materials possessed non-degenerate spin-up and spin-down states, allowing selective spin control. These are essential in devices such as spin filters and spin valves for data storage and memory applications in spintronics \cite{Xingxing2024APS,Zahir2020Elsevier,Kaustav2023AdvMat}. Furthermore, such bipolar magnetic semiconductors can be used in spin filters, spin valves, and potentially in energy-efficient data storage and quantum computing applications due to their ability to maintain distinct spin channels under external manipulations such as strain or doping \cite{Jinlong2016NationalScienceReview}.

\begin{table*}[ht]
\centering
\small 
\caption{Electronic properties of pristine TMDCs: Bandgap, Spin-polarized Bandgap, and Valley Splitting Energy}
\label{table:merged_TMDCs}
\resizebox{\textwidth}{!}{ 
\begin{tabular}{lccccccc}
\hline
\textbf{Crystals} & \textbf{Bandgap (DFT+U)} & \textbf{Bandgap (HSE06)} & \textbf{Spin-up Bandgap} & \textbf{Spin-down Bandgap} & \textbf{Valley splitting energy} & \textbf{Valley splitting energy} \\
 & \textbf{(eV)} & \textbf{(eV)} & \textbf{(eV)} & \textbf{(eV)} & \textbf{Valence Band (meV)} & \textbf{Conduction Band (meV)} \\

\hline
NbS$_2$ & 0.5818 & 0.2188 & 0.8227 & 1.5359 & -176.2 & 3.8 \\
NbSe$_2$ & 0.7911 & 0.4645 & 1.0254 & 1.3915 & -251.6 & 18.6 \\
NbTe$_2$ & 0.7700 & 0.3347 & 0.8556 & 0.9279 & -351.2 & 26.2 \\
TaSe$_2$ & 0.3977 & 0.1025 & 1.0759 & 1.3269 & 539.4 & 14.4 \\
TaTe$_2$ & 0.1947 & 0.9913 & 0.9159 & 0.7291 & 541.0 & 18.9 \\
\hline
\end{tabular}
}
\end{table*}

A substantial change in band structure was observed with the association of SOC, which breaks the degeneracy, and an energy difference was exhibited at K and $\mathrm{K}^\prime$ points. The energy splitting is defined as valley polarization. A large valley polarization was achieved at the valence band without applying any external factors. The valley splitting energy was calculated by,
\begin{equation}
\Delta E_{\text{valence}} = E_{\text{K(valence)}} - E_{\text{$\mathrm{K}^\prime$(valence)}}, and
\end{equation}
\vspace{-1.5em}
\begin{equation}
\Delta E_{\text{conduction}} = E_{\text{K(conduction)}} - E_{\text{$\mathrm{K}^\prime$(conduction)}}.
\end{equation}
The valley splitting energy was maximum for TaTe$_2$, which was 541.0 meV between K and $\mathrm{K}^\prime$. To achieve the same valley splitting, an external magnetic field of around 360 Tesla would be required. To be viable for valleytronic applications, a material must demonstrate broken valley degeneracy and exhibit a sufficiently large valley splitting to remain stable against thermal fluctuations. Studies suggest that a valley splitting of approximately 100 meV is necessary for stable operation at room temperature, ensuring resilience against thermal noise and enabling reliable manipulation of valley states for practical device implementations \cite{Xiao2018Wiley}. 

We observed a significant valley splitting in these monolayers. The valley splitting energies at the valence band in the TMDCs were 176.2, 251.6, 351.2, 539.4, and 541.0 meV, respectively. This supported nonvolatile valley polarization at room temperature. This also enables valley states in applications like quantum computing and valley-based logic, where stable and controllable states are crucial for processing and storing information \cite{Xiaodong2016NatureReview,Kaustav2023AdvMat}. The strong SOC in these materials contributed to this large valley splitting, making it comparable to advanced valleytronic materials such as MoS$_2$, which also achieved valley polarization through SOC effects \cite{Xiaodong2012Nature}.

For the conduction band, the maximum valley splitting was achieved for NbTe$_2$, which was 26.2 meV. The valley splitting energy was less than the ambient temperature noise; hence, this splitting is not discussed further. The valley splitting energy of the pristine crystals without applying any external methods is illustrated in Table \ref{table:merged_TMDCs}. Moreover, the band structure incorporating SOC effect is given in Fig.\,\ref{fig: band_structure_TMDCs}.
The association of SOC kept the band structure spin-polarized with a finite bandgap for NbS$_2$, NbSe$_2$, and NbTe$_2$. These materials were BMS after association of SOC. As a result, the spin filtering character prevailed for them. However, the effect of SOC raised the valence band above the Fermi level for TaSe$_2$ and TaTe$_2$, which made these materials half-metal ferromagnetic materials. These materials can produce 100\% spin polarization and be used for pure spin generation.

The investigation further showed that no spontaneous valley polarization was noticed in the AFM configuration of the pristine crystals, where there is reversal symmetry. This result indicated that the spontaneous valley polarization in these materials was due to ferromagnetism, which causes the breaking of TRS, and this also limits the anomalous valley Hall effect and valleytronic applications. The AFM band structure is illustrated in Fig.\,\ref{fig: manipulation} and the Supplementary Material.

To investigate the materials further, we reversed the magnetic moment of the transition atoms, which gave rise to the reversal of spin polarization at the valence and conduction bands. With the reversal of spin polarization, the valley polarization was also reversed. This gave a means to manipulate the valley polarization at K and $\mathrm{K}^\prime$ valleys. Due to magnetic moment reversal, valley polarization was reversed for NbS$_2$, NbSe$_2$, and NbTe$_2$. The magnitude of valley polarization exhibited negligible variation. However, the magnitude of valley polarization was also changed for TaSe$_2$ and TaTe$_2$ along with the reversal of valley polarization. For TaSe$_2$, the valley polarization was increased to 546.7 meV; for TaTe$_2$, the valley splitting energy was increased to 574.3 meV. The reversal of valley polarization and the magnitude of valley splitting energy are elucidated in Fig.\,\ref{fig: manipulation} and the Supplementary Material.

\subsection{Orbital Projected Band and DOS}
To get a nuanced view of the electronic properties, we calculated the orbital contribution of the materials at the valence band maximum and conduction band minimum. The orbital contribution at K and $\mathrm{K}^\prime$ valleys was also calculated to get a deep view of the character of the valleys, which showed valley polarization. The electronic structure analysis of TMDCs revealed significant contributions from \( p \) orbitals from chalcogens (C) and \( d \) orbitals from transition metals (TM).

The orbital contribution at the valence band's K and $\mathrm{K}^\prime$ points showed an interesting view. As K and $\mathrm{K}^\prime$ points were degenerate without SOC, the orbital contributions at K and $\mathrm{K}^\prime$ were identical. For NbS$_2$, the K and the $\mathrm{K}^\prime$ points at the valence band were contributed by \(d_{x^2-y^2}\), \(d_{xy}\), \(d_{yz}\), and \(d_{zx}\) orbitals of the Nb atom. The contributions were 21.4\%, 21.4\%, 6.1\%, and 6.1\%, respectively. The Se atom's \(p_x\) orbital had the majority contribution of 38.3\% at the K and $\mathrm{K}^\prime$ points at the valence band. In NbSe$_2$, K and $\mathrm{K}^\prime$ points at BZ were contributed by \(d_{x^2-y^2}\), and \(d_{xy}\) orbitals of the Nb atom, which were 38.5\% for both of these orbitals. The Se atom's \(p_x\) and \(p_y\) orbitals had the same contribution of 11.5\% at those points. A different behavior was seen for NbTe$_2$, TaSe$_2$, and TaTe$_2$. The K and $\mathrm{K}^\prime$ points were dominantly contributed by \(p_z\) the orbitals of the chalcogen atom. The contributions were 62.6\%, 73.1\%, and 61.1\% for \(p_z\) orbitals in those TMDCs, respectively. Along with \(p_z\), contribution from \(d_{zx}\), and \(d_{yz}\) orbitals of transition metal atoms were exhibited. The contribution for both orbitals was the same in magnitude. The contribution from both of them was 18.7\%, 13.5\%, and 16\% in NbTe$_2$, TaSe$_2$, and TaTe$_2$, respectively. The orbital contribution at K and $\mathrm{K}^\prime$ valleys of the valence band is presented in Table \ref{Table:orbital_contributions_valley}.
\begin{figure*}[ht]
    \centering
    \includegraphics[width=\textwidth]{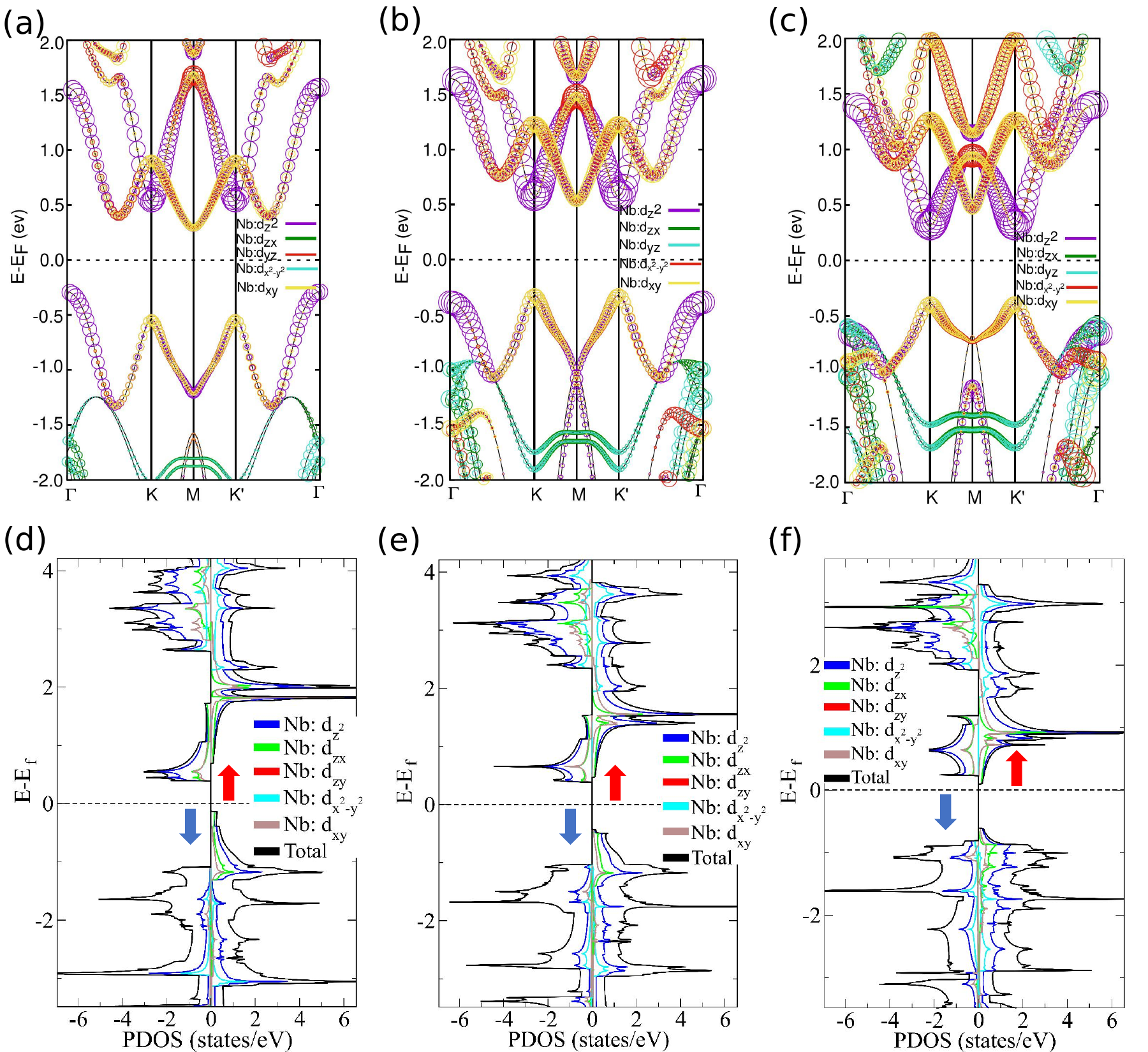} 
    \caption{The \(d\) orbitals projected E-k diagram without SOC for different TMDCs -- (a) NbS$_2$, (b) NbSe$_2$,  (c) NbTe$_2$. In the band structure, the size of the circles represents the amount of contribution. The illustration of the partial density of states of \(d\) orbitals for different TMDCs -- (d) NbS$_2$, (e) NbSe$_2$, (f) NbTe$_2$. Spin-up and spin-down states are indicated by the red up arrow and the blue down arrow, respectively.}
    \label{fig: fatband}
\end{figure*}

At the VBM of NbS\(_2\), the Nb \( d_{z^2} \) orbital was the dominant contributor, accounting for approximately 60.3\%, while the S atom's \( p_z \) orbital contributed around 37.7\%. In NbSe\(_2\), the Nb VBM was contributed by three degenerate \(d_{xy}\) , \(d_{x^2-y^2}\), and \(d_{zx}\) orbitals, with a contribution of 38.1\% from each orbital, with the Se atoms \( p_x \) and \(p_y\) orbitals contributing 11.4\% each. The same trend was seen for NbTe\(_2\), the VBM was dominated by three degenerate \(d_{xy}\) , \(d_{x^2-y^2}\), and \(d_{zx}\) orbitals, \( d_{z^2} \) with orbitals contributing 39.1\% from each. Te atoms \( p_x \) and \(p_y\) orbitals showed similar behavior like NbSe$_2$, contributing around 10.5\%. For TaSe$_2$, the VBM was mostly contributed by \(d_{z^2}\) the Ta atom. The contribution of this orbital was 67\%. The chalcogen atom's \(p_z\) orbital also showed a mentionable contribution in the VBM by contributing 30\%. TaTe$_2$ VBM showed the same behavior as NbSe$_2$ and NbTe$_2$. It contained three degenerate \(d_{xy}\) , \(d_{x^2-y^2}\), and \(d_{zx}\) orbitals. Each of the orbitals contributed 39.1\% in the VBM; 10.9\% was contributed by both \( p_x \) and \(p_y\) of the Te atom. These results underscored the predominant role of chalcogen \( p \) orbitals and transition metal \( d \) orbitals in shaping the VBM characteristics across these materials. 
\begin{table}[H] 
    \centering
    \caption{Orbital contributions in \% at K and $\mathrm{K}^\prime$ of valence band for different materials }
    \label{Table:orbital_contributions_valley}
    \scriptsize 
    \setlength{\tabcolsep}{2pt} 
    \begin{tabular}{lccccccc}
        \toprule
        \textbf{Materials} & \multicolumn{7}{c}{\textbf{K and $\mathrm{K}^\prime$}} \\
        \cmidrule(lr){2-8}
        & \textbf{TM: d$_{x^2-y^2}$} & \textbf{TM: d$_{xy}$} & \textbf{TM: d$_{yz}$} & \textbf{TM: d$_{zx}$} & \textbf{C: p$_{z}$} & \textbf{C: p$_{y}$} & \textbf{C: p$_{x}$}  \\
        \midrule
        NbS$_2$ & 21.4 & 21.4 & 6.1 & 6.1 & 0 & 6.7 & 38.3  \\
        NbSe$_2$ & 38.5 & 38.5 & 0 & 0 & 0 & 11.5 & 11.5 \\
        NbTe$_2$ & 0 & 0 & 18.7 & 18.7 & 62.6 & 0 & 0  \\
        TaSe$_2$ & 0 & 0 & 13.5 & 13.5 & 73.1 & 0 & 0  \\
        TaTe$_2$ & 0 & 0 & 16 & 16 & 61.1 & 0 & 0 \\
        \bottomrule
    \end{tabular}
\end{table}

\begin{table*}[b] 
    \centering
    \caption{Orbital contributions in \% for different materials at VBM and CBM}
    \label{Table:orbital_contributions}
    \scriptsize 
    \begin{tabular}{lccccccc cccccc}
        \toprule
        \textbf{Materials} & \multicolumn{7}{c}{\textbf{Valence Band Maximum (VBM)}} & \multicolumn{6}{c}{\textbf{Conduction Band Minimum (CBM)}} \\
        \cmidrule(lr){2-8} \cmidrule(lr){9-14}
        & \textbf{TM: d$_{x^2-y^2}$} & \textbf{TM: d$_{xy}$} & \textbf{TM: d$_{z^2}$} & \textbf{TM: d$_{zx}$} & \textbf{C: p$_{z}$} & \textbf{C: p$_{y}$} & \textbf{C: p$_{x}$} 
        & \textbf{TM: d$_{x^2-y^2}$} & \textbf{TM: d$_{xy}$} & \textbf{TM: d$_{z^2}$} & \textbf{TM: d$_{zx}$} & \textbf{C: p$_{z}$} & \textbf{C: p$_{x}$} \\
        \midrule
        NbS$_2$ & 0 & 0 & 60.3 & 0 & 37.7 & 0 & 0 & 8.9 & 26.7 & 28.3 & 26.7 & 17.1 & 2.8 \\
        NbSe$_2$ & 38.1 & 38.1 & 0 & 38.1 & 0 & 11.4 & 11.4 & 9.8 & 29.3 & 27.8 & 29.3 & 13.2 & 2.8 \\
        NbTe$_2$ & 39.1 & 39.1 & 0 & 39.1 & 0 & 10.5 & 10.5 & 0 & 0 & 82.6 & 0 & 0 & 4.1 \\
        TaSe$_2$ & 0 & 0 & 67.0 & 0 & 30.0 & 0 & 0 & 11.1 & 33.2 & 32.7 & 33.2 & 21.0 & 0.9 \\
        TaTe$_2$ & 39.1 & 39.1 & 0 & 39.1 & 0 & 10.9 & 10.9 & 11.5 & 34.5 & 32.6 & 34.5 & 20.4 & 0.5 \\
        \bottomrule
    \end{tabular}
\end{table*}
The CBM analysis also highlighted the dominant influence of transition metal \( d \) orbitals across the materials, with smaller contributions from the \( s \) and \( p \) orbitals of chalcogens. For NbS\(_2\), the CBM was consisted of \(d_{x^2-y^2}\), \(d_{xy}\), \(d_{z^2}\), and \(d_{zx}\). They had contributions of 8.9\%, 26.7\%, 28.3\%, and 26.7\%, respectively. \(p_z\) orbital of the S atom was also mentionable with a contribution of 17.1\%. The same behavior was seen for the CBM of NbSe$_2$. \(d_{x^2-y^2}\),\(d_{xy}\), \(d_{z^2}\), and \(d_{zx}\) had contributions of 9.8\%, 29.3\%, 27.8\%, and 29.3\%, respectively. The contribution from  \(p_z\) orbital of the Se atom was around 13.2\%. A different character was shown for the CBM of NbTe$_2$. \(d_{x^2-y^2}\),\(d_{xy}\), and \(d_{zx}\) orbitals had no contribution in CBM. CBM was contributed by \(d_{z^2}\) which was about 82.6\%. The contribution from chalcogens was also negligible. However, TaSe$_2$ followed the same trend as NbS$_2$ and NbSe$_2$. \(d_{x^2-y^2}\),\(d_{xy}\), \(d_{z^2}\), and \(d_{zx}\) orbitals showed a contribution of 11.1\%, 33.2\%, 32.7\%, and 33.2\%, respectively, and \(p_z\) orbitals of the Se atom contributed 21\% in the CBM. The contribution from  \(d_{x^2-y^2}\),\(d_{xy}\), \(d_{z^2}\), and \(d_{zx}\) orbitals in TaTe$_2$ were 11.5\%, 34.5\%, 32.6\%, and 34.5\%, respectively. Moreover, 20.4\% was contributed from \(p_z\) orbitals of Te atoms. These findings revealed the importance of transition metal \( d \) orbitals in defining the CBM across these TMDCs, with chalcogen \( p \) orbitals playing a supplementary role. The orbital contributions at VBM and CBM are illustrated in Table \ref{Table:orbital_contributions}.

The \(d\) orbital projected band structure and the partial density of states (PDOS) of \(d\) orbitals are presented in Fig.\,\ref{fig: fatband} and the Supplementary Material, showing the contribution in a more elaborated view. The provided figure of \(d\) orbital projected band structure and PDOS of \(d\) orbitals in Fig.\,\ref{fig: fatband} comprehensively illustrates the \(d\) orbital projected electronic band structures and PDOS for various TMDCs calculated without considering SOC. The band structure diagrams in Figs.\,\ref{fig: fatband}(a–c) reveal the contribution of different \(d\) orbitals (\(d_{z^2}\), \(d_{x^2-y^2}\), \(d_{xy}\), \(d_{xz}\), and \(d_{yz}\)) to the electronic states across the BZ. The size of the circles in these diagrams represents the relative weight of specific orbitals at various energy levels and k points. In the vicinity of the Fermi level (\(E - E_F = 0\)), notable contributions from the \(d_{z^2}\) and \(d_{xy}\), and \(d_{x^2-y^2}\) orbitals were observed at the valence band, particularly for NbS\(_2\), NbSe\(_2\), and TaSe$_2$, indicating that this orbital played a significant role in defining the electronic properties of these materials. Comparatively, in NbTe\(_2\), and TaTe$_2$ near the Fermi level at the valence band, \(d_{xy}\), and \(d_{x^2-y^2}\) orbitals were more important. At the conduction band near the Fermi level, \(d_{xy}\) and \(d_{x^2-y^2}\) were more important for NbS$_2$, TaSe$_2$, and TaTe$_2$. For NbSe$_2$, the closet point of the conduction band was dominated by \(d_{z^2}\) and \(d_{xy}\), and \(d_{x^2-y^2}\) orbitals. On the other hand, NbTe$_2$ \(d_{z^2}\) was dominant at the nearest point from the Fermi level. 
\begin{figure*}[H]
    \centering
    \includegraphics[width=\textwidth]{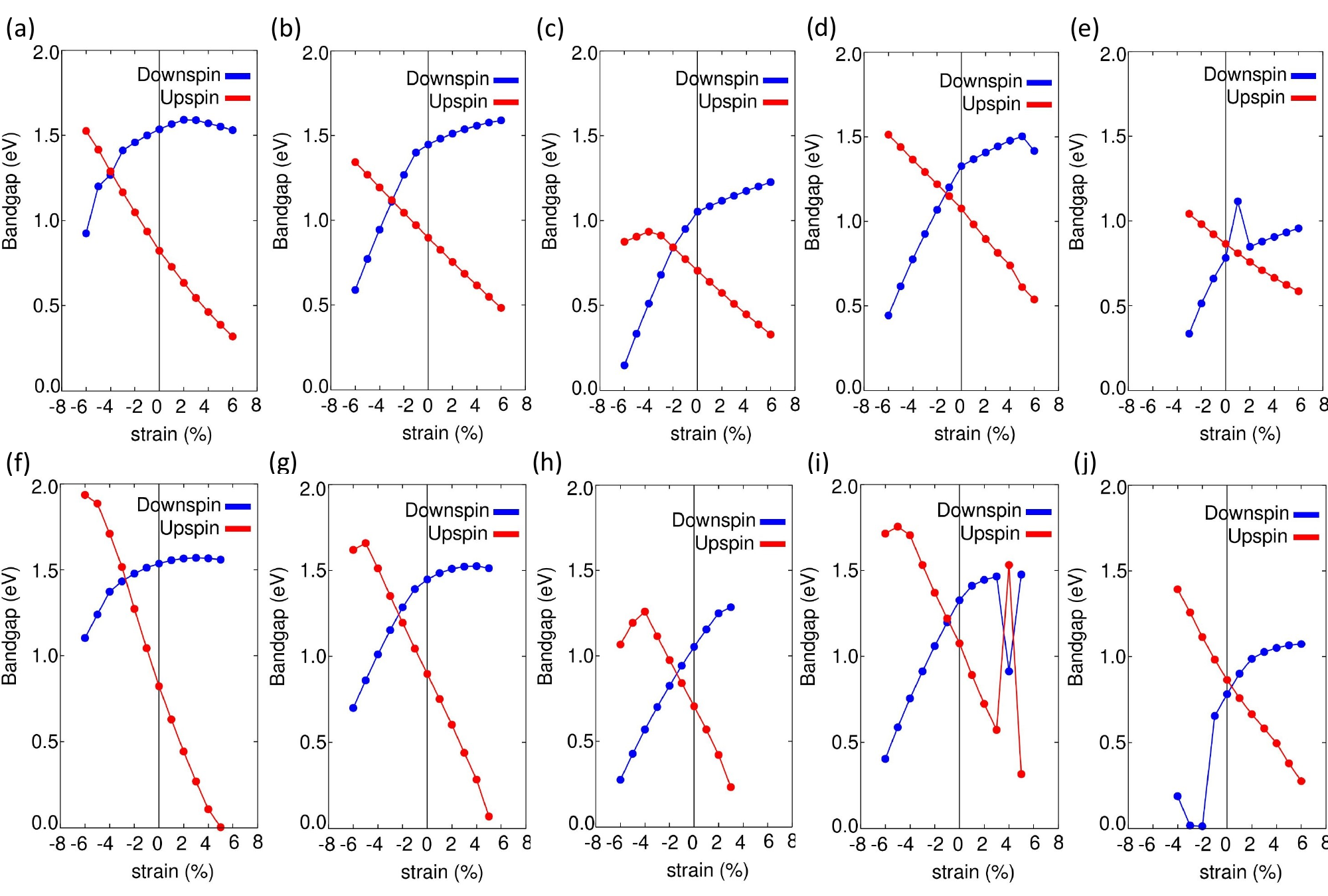} 
    \caption{Effect of uniaxial strain on spin-polarized bandgap for different TMDCs: (a) NbS$_2$, (b) NbSe$_2$,  (c) NbTe$_2$, (d) TaSe$_2$, (e) TaTe$_2$. The spin-down and spin-up bandgap are distinguished by blue and red lines. Effect of biaxial strain on spin-polarized bandgap for different TMDCs: (f) NbS$_2$, (g) NbSe$_2$,  (h) NbTe$_2$, (i) TaSe$_2$, (j) TaTe$_2$. Spin-down and spin-up bandgap are represented by blue and red lines, respectively. Positive strain refers to tensile strain, and negative strain refers to compressive strain. }
    \label{fig: bandgap_spin}
\end{figure*}
The PDOS plots in Fig.\,\ref{fig: fatband}(d–f) complement the band structure analysis by quantifying the DOS associated with individual \(d\) orbitals. In the pristine TMDCs, the \(d_{z^2}\) orbital exhibited a pronounced peak of DOS near the Fermi level at the valence band, which was contributed by the spin-up channel electrons. At the conduction band, the spin-down channel electron contributed to the DOS near the Fermi level, which was consistent with the electronic properties of BMS materials. The \(d_{z^2}\) orbital also exhibited a dominant DOS in the conduction band. The \(d_{x^2-y^2}\) and \(d_{xy}\) orbitals, while present, contributed less significantly. The other orbitals, \(d_{xz}\), and \(d_{yz}\) had negligible contribution at the DOS. 

The analysis of these data collectively underscored the importance of \(d\) orbital contributions in defining the electronic properties of TMDCs. The \(d_{z^2}\) and \(d_{xy}\) orbitals were found to play a dominant role across all compounds, particularly near the Fermi level, where they largely dictated the semiconducting behavior of the materials. Furthermore, the degree of orbital hybridization and band dispersion varied systematically across the chalcogenide series, indicating that the electronic properties can be tuned by doping or substituting chalcogen atoms or transition metals. These insights are pivotal for understanding the fundamental electronic behavior of TMDCs and their potential applications in electronic, optoelectronic, and superconducting devices. The projected band structure is provided in Fig.\,\ref{fig: fatband}.

\subsection{Effect of Uniaxial and Biaxial Strain}

\begin{figure*}[H]
    \centering
    \includegraphics[width=\textwidth]{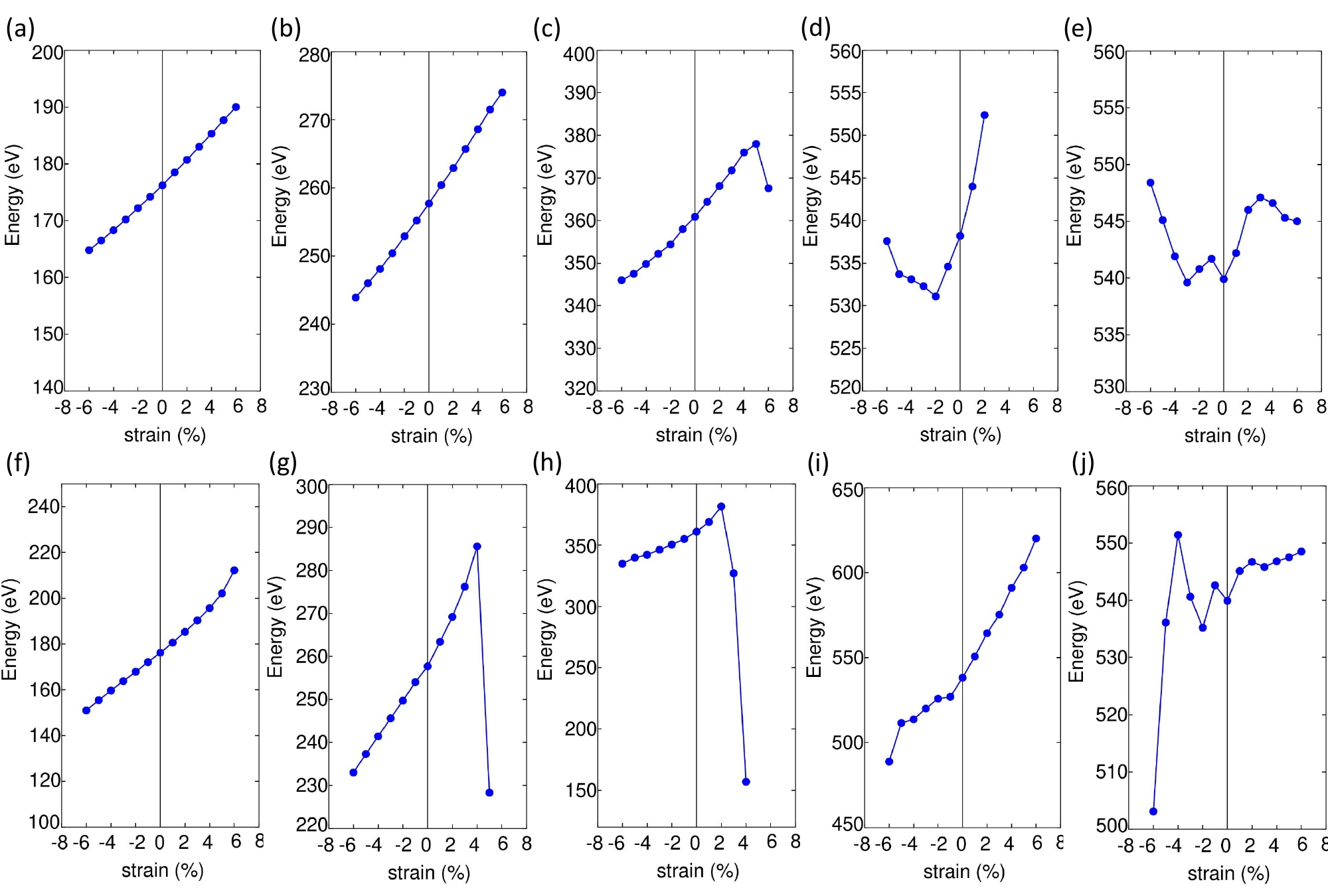} 
    \caption{ The effect of uniaxial strain in valley polarization is shown for different TMDCs: (a) NbS$_2$, (b) NbSe$_2$, (c) NbTe$_2$, (d) TaSe$_2$, (e) TaTe$_2$. The effect of biaxial strain on valley polarization is shown for different TMDCs: (f) NbS$_2$, (g) NbSe$_2$, (h) NbTe$_2$, (i) TaSe$_2$, (j) TaTe$_2$. Positive strain refers to tensile strain, and negative strain refers to compressive strain. }
    \label{fig: valley_polarization_strain}
\end{figure*}

The valley splitting can be enhanced through external modifications, such as strain or atomic vacancies \cite{Yan2024RSC}. Uniaxial and biaxial strains were applied to investigate the behavior of the materials under strain. The strain was applied by the formula
\begin{equation}
\text{Strain} = \frac{a' - a}{a} \times 100\%
\end{equation}
Here,  $a'$ was the strained lattice parameter, and $a$ is the relaxed lattice parameter. The strain impacted the bandgap, spin-up bandgap, spin-down bandgap, and valley polarization. For both uniaxial and biaxial strain in NbS$_2$ with increasing tensile strain, the spin-up band of the conduction band got closer to the Fermi level, narrowing the spin-up bandgap. The spin-down bandgap remained nearly constant with tensile strain. With increasing compressive strain, the spin-up bandgap increased, and the spin-down bandgap decreased as the spin-down band of the conduction band got close to the Fermi level. The exact impact was noticed for all other materials. With increasing tensile strain, the spin-up bandgap decreased, and the spin-down bandgap increased, and for increasing compressive strain, the spin-up bandgap increased, and the spin-down bandgap decreased. This trend was the same for uniaxial and biaxial strains. However, under 4\% compressive uniaxial strain and without SOC, TaTe$_2$ became a half-metal ferromagnet. For biaxial strain, NbS$_2$, TaSe$_2$, and TaTe$_2$ became half-metal ferromagnets at 5\% tensile strain, 6\% compressive strain, and 2\% compressive strain, respectively, showing a great potential to be used to generate pure spin-polarized electrons. However, for 5\% compressive strain, TaTe$_2$ turned into a full metallic ferromagnet. The impact of uniaxial and biaxial strains on spin-polarized bandgap is illustrated in Fig.\,\ref{fig: bandgap_spin}.

Moreover, the impact of strain on the valley polarization was noticeable. With the increasing tensile strain, the magnitude of the valley polarization in the valence band was increased. The valley polarization at the valence band decreased with the increasing compressive strain. The trend was the same for both uniaxial and biaxial strains. The valley polarization at the valence band for NbS$_2$, NbSe$_2$, NbTe$_2$, and TaSe$_2$ more or less showed a linear trend with uniaxial and biaxial strain. For TaTe$_2$, the impact of uniaxial and biaxial strain was elusive and showed no linear trend. Moreover, the conduction band experienced variation with the application of strain. As the valley polarization was smaller than 100 meV, the discussion was dropped because it could not surpass the ambient thermal noise. The magnitude of valley polarization could also be manipulated by uniaxial and biaxial strain. Fig.\,\ref{fig: valley_polarization_strain} summarizes the effect of uniaxial and biaxial strain in valley polarization.

\subsection{Berry Curvature and Anomalous Hall Effect}
\begin{figure*}[H]
    \centering
    \includegraphics[width=\textwidth]{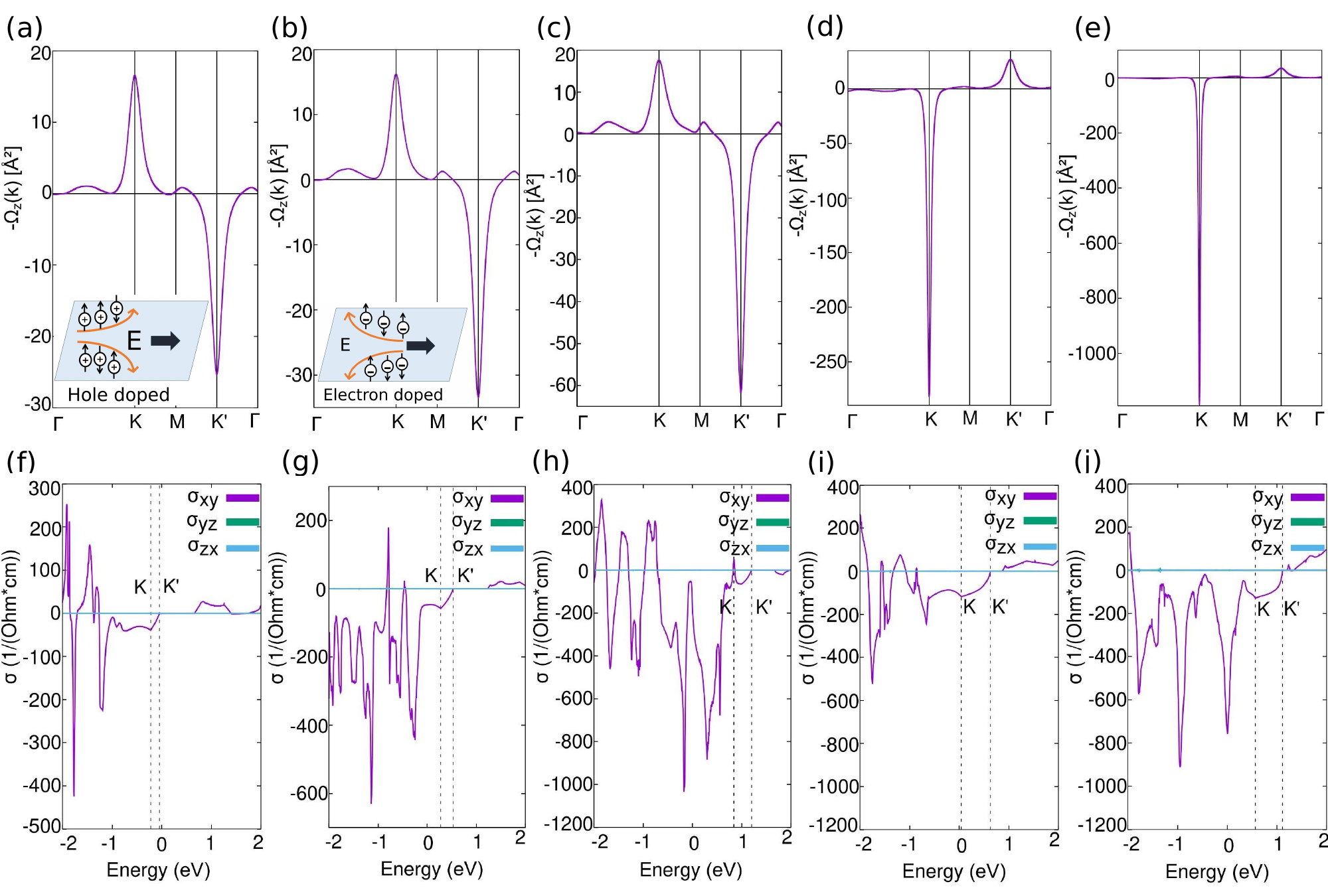} 
    \caption{Berry curvature in the valence band in different TMDCs: (a) NbS$_2$, (b) NbSe$_2$, (c) NbTe$_2$, (d) TaSe$_2$ (4\% strain), (e) TaTe$_2$ (5\% strain). Anomalous Hall conductivity in the valence band for different TMDCs is given: (f) NbS$_2$, (g) NbSe$_2$, (h) NbTe$_2$, (i) TaSe$_2$ (4\% strain), (j) TaTe$_2$ (5\% strain). Insets illustrate the anomalous Hall effect for (a) hole-doped and (b) electron-doped systems, showing valley-dependent transverse motion of carriers under an electric field, $E$, driven by valley-specific Berry curvature.} 
    \label{fig: berry_curvature}
\end{figure*}

Berry curvature is crucial in modern condensed matter physics, particularly in materials with broken spatial inversion or broken TRS. It acts as an effective magnetic field in momentum space for the Bloch electrons \cite{Qian1996APS}. The Berry curvature, denoted as \( \Omega(\mathbf{k}) \), originates from the geometrical properties of the wave functions in the crystal’s BZ. For a given electronic band \( n \), the Berry curvature at a point \( \mathbf{k} \) in the BZ was defined by,

\begin{equation}
\Omega_n(\mathbf{k}) = -2\,\text{Im}\Bigl(
  \sum_{m \neq n}
    \frac{\langle u_{n\mathbf{k}} | \hat{v}_x | u_{m\mathbf{k}} \rangle
          \langle u_{m\mathbf{k}} | \hat{v}_y | u_{n\mathbf{k}} \rangle}
         {(E_{n\mathbf{k}} - E_{m\mathbf{k}})^2}
\Bigr).
\end{equation}

Where \( | u_{n\mathbf{k}} \rangle \) and \( | u_{m\mathbf{k}} \rangle \) were the periodic parts of the Bloch wavefunctions for bands \( n \) and \( m \), \( \hat{v}_x \) and \( \hat{v}_y \) were the velocity operators in the \( x \)- and \( y \)-directions, and \( E_{n\mathbf{k}} \) and \( E_{m\mathbf{k}} \) were the energy eigenvalues. This expression highlights that the Berry curvature arises from the interband matrix elements between occupied and unoccupied bands. The magnitude of Berry curvature was opposite at K and $\mathrm{K}^\prime$ points in the BZ due to the absence of TRS \cite{Qian2010APS,Zongwen2019NanoResearch}. This is why, in the presence of an in-plane electric field, the Bloch electrons of K and $\mathrm{K}^\prime$ valley gain a transverse velocity associated with Berry curvature called anomalous velocity \cite{nagaosa2010APS}. This gave rise to a transverse current and Hall conductance \cite{Jiwoong2014AmericanAssociation}. In systems with broken inversion symmetry and broken time reversal symmertry, Berry curvature exhibits valley-contrasting behavior, where it has opposite signs and unequal magnitude in the K and $\mathrm{K}^\prime$ valleys, leading to valley-dependent phenomena like the valley Hall effect \cite{Qian1996APS}. 

At K and $\mathrm{K}^\prime$ points in the valence band, the Berry curvature exhibited opposite signs with unequal magnitudes in NbS$_2$, NbSe$_2$, NbTe$_2$, TaSe$_2$, and TaTe$_2$. These materials showed an anomalous behavior in Berry curvature. NbS$_2$, NbSe$_2$, and NbTe$_2$ showed opposite Berry curvature at K and $\mathrm{K}^\prime$ points in BZ without external factors. For TaSe$_2$ and TaTe$2$, by applying 4\% and 5\% strain, respectively, Berry curvature at K and $\mathrm{K}^\prime$ was achieved at VBM in BZ. Due to opposite Berry curvature, electrons at K valley and $\mathrm{K}^\prime$ valley will experience opposite transverse velocity with the application of an in-plane electric field. This phenomenon was confirmed by the anomalous Hall conductivity. The anomalous Hall effect (AHE) is a direct consequence of non-zero Berry curvature. Unlike the conventional Hall effect, which requires an external magnetic field, the AHE arises from the intrinsic Berry curvature, which deflects charge carriers transversely to an applied electric field. The anomalous velocity of charge carriers due to Berry curvature \( \Omega(\mathbf{k}) \) is given by
\begin{equation}
\mathbf{v}_\perp = \frac{e}{\hbar} \, \mathbf{E} \times \Omega(\mathbf{k})
\end{equation}
Where \( \mathbf{E} \) is the applied electric field. This anomalous velocity leads to a transverse current, manifesting as the AHE without an external magnetic field. The AHC, \( \sigma_{xy} \), can be computed by integrating the Berry curvature over all occupied states in the BZ
\begin{equation}
\sigma_{xy} = \frac{e^2}{\hbar} \int_{\text{BZ}} \frac{d^2k}{(2\pi)^2} \, f(\mathbf{k}) \, \Omega(\mathbf{k})
\end{equation}
where \( f(\mathbf{k}) \) is the Fermi-Dirac distribution function, accounting for the occupation of electronic states. This formula indicates that AHC depends on the Berry curvature distribution across the BZ and the position of the Fermi level.
Berry curvature and AHE are particularly significant in systems with strong SOC and broken inversion symmetry, such as TMDCs. In these materials, the Berry curvature in the K and $\mathrm{K}^\prime$ valleys was large and oppositely signed, resulting in valley-contrasting AHE \cite{Zongwen2019NanoResearch}. This valley-selective Berry curvature and anomalous Hall effect enable applications in valleytronics, where information can be stored and manipulated based on the valley index. The materials showed AHC at VBM in K and $\mathrm{K}^\prime$ points. At K point in the BZ, the valley Hall conductivity was seen and at $\mathrm{K}^\prime$ the conductivity is zero, suggesting that with the application of in-place electric field carriers, from K and $\mathrm{K}^\prime$ valleys will move towards two edges of materials and give rise to anomalous Hall voltage. For NbS$_2$, NbSe$_2$, and NbTe$_2$ spontaneous AHC were seen, whereas for TaSe$_2$ and TaTe$_2$, like in Berry curvature, an application of 4\% and 5\% biaxial strain was applied to calculate the AHC.
The Berry curvature and the AHC is illustrated in Fig.\,\ref{fig: berry_curvature}.

\section{Conclusion}
We extensively explored the electronic and magnetic characteristics of a novel class of FV materials, namely NbX$_2$ and TaX$_2$ (X = S, Se, Te). Through first-principles calculations, this investigation showed that these TMDCs possess stable FM ordering. That led to giant and tunable spontaneous valley polarization at the valence band without any external manipulation, which was 176.2, 251.6, 351.2, 539.4, and 541 meV for NbS$_2$, NbSe$_2$, NbTe$_2$, TaSe$_2$, and TaTe$_2$, respectively. This valley polarization was large enough to be sustained at room temperature and even in harsh conditions with high temperatures. Additionally, as BMS, they allowed for fully spin-polarized currents with reversible spin orientation, which can be controlled by gate voltage. NbS$_2$, TaSe$_2$, and TaTe$_2$ became half-metal ferromagnets with the application of 5\%, 6\%, and 2\% biaxial strain, respectively, which showed the potential to generate a fully spin-polarized electron. This unique feature supports the development of efficient valleytronic and spintronic devices. Besides this, SOC and lack of inversion symmetry in these materials resulted in pronounced valley splitting at the K and $\mathrm{K}^\prime$ points within the BZ, creating a non-degenerate energy configuration that led to the anomalous Hall effect. Additionally, the ability to modulate valley polarization and spin-polarized bandgap through external strain and electric fields further positions these dichalcogenides as strong candidates for non-volatile memory, logic devices, and spin-filtering applications. The Nb and Ta-based dichalcogenides examined here represent a promising foundation for future valleytronic innovations, providing intrinsic valley-polarized current control without external magnetic fields, in line with the requirements of emerging data storage and processing technologies. This study lays critical groundwork for understanding these materials and highlights their potential to drive progress in spintronics and valleytronics.
\FloatBarrier  

\section*{Associated Content}
\noindent \textbf{Supplementary Material:} \\
Supplementary Material is available free of charge at \url{url}. This includes spin polarization analysis, band diagrams of pristine NbS$_2$ and NbSe$_2$, spin reversal band diagram and AFM band diagram, and orbital projected band structure and DOS analysis for various orbitals in the structures.

\section*{Declaration of Competing Interest}
The authors declare that they have no known competing financial interests or personal relationships that could have appeared to influence the work reported in this paper.

\section*{Data Availability Statement}
{The data supporting the findings presented in this paper are not currently available to the public, but they may be obtained from the authors upon reasonable request.} 

\balance

\printcredits  

\bibliography{references}

\end{document}